\documentclass[aps,prl,twocolumn, preprintnumbers]{revtex4}
\usepackage{graphicx}
\usepackage{bm}

\begin{document}

\preprint{KEK-TH-980}
\title{Decoupling Supersymmetry/Higgs without fine-tuning }

\author{Naoyuki Haba} 
\affiliation{Institute of Theoretical Physics, 
University of Tokushima, Tokushima 770-8502, Japan} 
\author{Nobuchika Okada}
\affiliation{Theory Division, KEK, Tsukuba, Ibaraki 305-0801, Japan}
%
%\date{\today}

\begin{abstract}
We propose a simple superpotential for the Higgs doublets, 
 where the electroweak symmetry is broken 
 at the supersymmetric level. 
We show that, 
 for a class of supersymmetry breaking scenarios, 
 the electroweak scale can be stable 
 even though the supersymmetry breaking scale 
 is much higher than it. 
Therefore, 
 all the superpartners and the Higgs bosons 
 can be decoupled from the electroweak scale, 
 nevertheless no fine-tuning is needed. 
We present a concrete model, 
 as an existence proof of such a model, 
 which generates the superpotential dynamically. 
According to supersymmetry breaking scenarios to be concerned, 
 various phenomenological applications of our model are possible. 
For example, based on our model, 
 the recently proposed ``split supersymmetry'' scenario 
 can be realized without fine-tuning. 
If the electroweak scale supersymmetry breaking 
 is taken into account, 
 our model provides a similar structure 
 to the recently proposed ``fat Higgs'' model 
 and the upper bound 
 on the lightest Higgs boson mass can be relaxed.

\end{abstract}

%\pacs{Valid PACS appear here}

\maketitle

%%%%%%%%%%%%%%%%%%% 
%    Main Body    %
%%%%%%%%%%%%%%%%%%%
Although the standard model 
 (with a simple extention  
 to incorporate neutrino masses and mixings) 
 is in good agreement with 
 almost of all the current experimental data, 
 the model suffers from many problems. 
The gauge hierarchy problem is the most serious one. 
Since, in quantum theories,  
 the Higgs boson mass is quadratically sensitive 
 to new physics scale, 
 an extreme fine-tuning is inevitable 
 in order to obtain the correct electroweak scale 
 if new physics scale lies at, for example, the Planck scale. 
In other words, the vacuum in the standard model 
 is not stable against quantum corrections. 

It is well known that this fine-tuning problem can be solved 
 by introducing supersymmetry (SUSY) \cite{SUSY-review}. 
In SUSY models, there is no quadratic divergence 
 of quantum corrections for the Higgs mass  
 by virtue of supersymmetry, 
 and hence the stability of the electroweak scale is ensured. 
However, none of superpartners have been observed yet, 
 the SUSY must be broken at low energies. 
Once the SUSY is broken, 
 the quadratic sensitivity returns 
 and the Higgs boson mass receives 
 quantum corrections of the SUSY breaking scale. 
Therefore, the SUSY breaking scale 
 should be of the electroweak scale or smaller. 
Otherwise, the fine-tuning problem returns 
 and the motivation of introduction of SUSY may fade away. 
Unfortunately, the lower bound on masses 
 of superpartners and the lightest Higgs boson 
 in the minimal supersymmetric standard model (MSSM) 
 is being raised by the current experiments. 
In the MSSM, we have already faced 
 the so-called ``little hierarchy problem'' 
 and a fine-tuning at the accuracy of about 1 \% 
 in the Higgs potential 
 is needed to obtain the correct scale of 
 the electroweak symmetry breaking. 
This fact may indicate 
 that we should admit the fine-tuning at this (or higher) level 
 or 
 the Higgs sector in the MSSM should be extended 
 so that the lightest Higgs boson can be heavier \cite{fat-Higgs} 
 or 
 the nature is fine-tuned intrinsically 
 and SUSY is nothing to do with 
 the gauge hierarchy problem \cite{Split-SUSY}. 

Here let us reconsider the fine-tuning problem. 
The quadratic ultraviolet sensitivity of 
 quantum corrections for the scalar mass 
 is inevitable in field theories 
 without SUSY or with broken SUSY. 
However, note that the following conditions are satisfied, 
 the problem can be avoided: 
 the Higgs boson is heavy enough to neglect 
 the quadratic quantum corrections, but 
 develops the electroweak scale VEV. 
Normally such a situation cannot be realized. 
This is because a usual Higgs potential in the standard model 
 contains only one mass parameter such that 
\begin{eqnarray} 
V(\phi) = \mu \phi^\dagger \phi + \lambda (\phi^\dagger \phi)^2,  
\end{eqnarray} 
where $ \phi $ is the Higgs doublet, 
 and $\mu <0 $ is the negative mass squared. 
The Higgs boson mass ($ \sqrt{2 |\mu|} $) 
 is found to be the same order of 
 or smaller than the scale of its VEV 
 ($ | \langle \phi \rangle | =  \sqrt{|\mu|/2 \lambda}$), 
 unless the dimensionless coupling $\lambda $ is taken 
 to be large beyond the perturbative regions. 

If the Higgs potential includes 
 two (or more) hierarchical mass parameters, 
 it may be possible to obtain a Higgs VEV 
 much smaller than the Higgs boson mass itself 
 through a combination among mass parameters. 
In neutrino physics, 
 there is a famous example, 
 namely the see-saw mechanism \cite{see-saw}, 
 which can effectively lead to the tiny neutrino mass scale 
 through a relation among hierarchical mass scales. 
If a similar mechanism works in a Higgs potential 
 the fine-tuning problem may be solved 
 (the same arguments are seen in Ref.~\cite{Atwood-etal}). 
In this paper, we introduce a supersymmetric model 
  which can realize such a situation. 

We propose a simple Higgs superpotential 
 with two mass parameters such that 
\begin{eqnarray} 
W= m H_u H_d 
 + \alpha^{-1} \frac{M^{3+2 \alpha}}{(H_u H_d)^{\alpha}} 
 \label{superpotential}
\end{eqnarray} 
where $H_u$ and $H_d$ are 
 the up and the down-type Higgs doublets, respectively, 
 $m$ and $M$ are mass parameters, 
 and $\alpha >0$ is a positive number. 
The first term is a mass term, 
 while the second term is the so-called runaway superpotential. 
Suppose that the hierarchy $ M \ll m$ in the following. 
Before discussing the origin of the superpotential, 
 let us first show consequences derived from it. 

The Higgs doublets obtain their VEVs through SUSY vacuum conditions 
 and the electroweak symmetry is broken. 
The F-flat condition leads to 
\begin{eqnarray} 
  \langle H_u^0 H_d^0 \rangle = M^2 
  \left( \frac{M}{m} \right)^{\frac{1}{\alpha +1}}
 \label{vev}
\end{eqnarray} 
where $H_u^0$ and $H_d^0$ are the electric charge neutral components 
 in each Higgs doublets, 
 and furthermore 
 $\langle H_u^0 \rangle = \langle H_d^0 \rangle $ is required 
 by the D-flat conditions. 
Note that since $ M \ll m$ 
 we can obtain the electroweak scale 
 much smaller than $M$ (and $m$) 
 through the similar relation to the see-saw mechanism. 
Analyzing the superpotential 
 and the (supersymmetric) Higgs potential 
 including the D-term potentials, 
 we can check the supersymmetric mass spectrum 
 consistent with the supersymmetric Higgs mechanism, 
 namely there are three would-be Nambu-Goldstone (NG) bosons, 
 one real and one complex Higgs bosons 
 and their superpartners, 
 which play a role of component fields 
 in the massive vector multiplets of Z and W bosons. 
Note that, in addition to them, 
 there exists one (neutral) chiral Higgs multiplet 
 with heavy mass $ 2 (\alpha +1) m$. 
This heavy Higgs boson parameterizes the direction 
 perpendicular to the F-flat direction of Eq.~(\ref{vev}), 
 on the other hand, 
 the F-flat direction itself is bounded 
 by only the D-flat conditions 
 and is parameterized by the light Higgs bosons 
 being the scalar components 
 in the massive vector multiplets. 

In order for the model to be realistic, 
 SUSY should be broken. 
After the SUSY is broken, 
 the soft SUSY breaking terms 
 are taken into account in the Higgs potential. 
Since the direction perpendicular to the F-flat condition 
 is bounded by the large mass scale $m$, 
 Eq.~(\ref{vev}) is approximately satisfied 
 at a potential minimum, namely 
 $ \langle H_u^0 \rangle \simeq v^2/ \langle H_d^0 \rangle $ 
 where $v^2= M^2 ( M/m )^{\frac{1}{\alpha+1}}$, 
 even in the presence of any soft SUSY breaking terms 
 smaller than $m$. 
On the other hand, 
 $ \langle H_u^0 \rangle $ itself 
 is in general sensitive to the soft SUSY breaking terms 
 larger than the electroweak scale, 
 since the F-flat direction is bounded by only the D-flat conditions. 
If $\langle H_u^0 \rangle $ remains in the electroweak scale, 
 in other words, 
 $\tan \beta = \langle H_u^0 \rangle/ \langle H_d^0 \rangle $ 
 remains to be of order one 
 even though the soft SUSY breaking terms are 
 much higher than the electroweak scale, 
 the fine-tuning problem can be solved. 

Now we analyze the Higgs potential 
 including the soft SUSY breaking terms. 
We parameterize them such that 
\begin{eqnarray} 
 V_{\mbox{soft}} &=& m_u^2 |H_u|^2 + m_d^2 |H_d|^2 
 \nonumber \\
  &+&  B_1  m H_u H_d + B_2 \frac{m v^{2(\alpha +1)}}{(H_u H_d)^\alpha}.  
\end{eqnarray} 
Let us consider some typical cases. 
Case (i): $|B_1| \sim |B_2|  \gg |m_u| \sim |m_d|$. 
This case is naturally realized in 
 the anomaly mediated supersymmetry breaking (AMSB) scenario 
 \cite{AMSB1, AMSB2} 
 with the sequestering ansatz \cite{AMSB1}, 
 where $|B_{1,2}| \sim m_{3/2}$ 
 and $|m_{u,d}| \sim 0.01 m_{3/2}$  
 ($m_{3/2}$ is the gravitino mass). 
The mass terms can be neglected as a good approximation. 
Considering a symmetry under $H_u \leftrightarrow H_d$,  
 we can easily solve the stationary conditions analytically 
 and find a potential minimum at 
 $\langle H_u^0 \rangle = \langle H_d^0 \rangle 
  = v (1+ {\cal O}(m_{3/2}/m)) $. 
Note that the VEV is slightly shifted from the value 
 in the SUSY limit due to the SUSY breaking effect. 
At this vacuum, 
 the light Higgs bosons in the SUSY limit 
 obtain their masses of order $m_{3/2}$ 
 through the soft SUSY breaking terms. 
Furthermore, their superpartner Higgsinos 
 also obtain masses of order $m_{3/2}$. 
This can be understood as follows. 
In the SUSY limit, there is no $\mu$ term 
 ($\mu$ parameter in the chargino and neutralino mass matrices), 
 since $\mu = 0 $ is obtained under the F-flat condition. 
After SUSY is broken, 
 the Higgs VEV is sifted by 
 $ v \times {\cal O}(m_{3/2}/m)$ from the value in the SUSY limit, 
 and thus the F-flat condition is no longer exactly satisfied. 
As a result, the $\mu$ term of order $m_{3/2}$ is generated. 
This is the same mechanism discussed in \cite{Kitano-Okada}. 
Therefore, the electroweak scale can be stable 
 almost independently of the soft SUSY breaking terms. 
We can raise the SUSY breaking scale without fine-tuning 
 so that all the superpartners and also Higgs bosons are decoupled. 
Case (ii): $|m_u| \sim  |m_d|  \gg |B_1| \sim |B_2|$. 
This case is naturally realized in 
 the gauge mediated SUSY breaking scenario \cite{GMSB}. 
We can neglect B-terms as a good approximation. 
As far as $|m_{u,d}| \ll m$, 
 the F-flat condition of Eq.~(\ref{vev}) 
 is almost satisfied, 
 and the Higgs potential can be approximately reduced 
 into the form,  
\begin{eqnarray} 
V \sim  m_u^2 |H_u|^2 + m_d^2 \frac{v^4}{|H_u|^2} 
      + \mbox{$D$-terms} .  
 \label{case 2}
\end{eqnarray}  
As discussed above, 
 the F-flat direction is bounded by only the D-term potential. 
Note that this situation is the same as 
 for squarks and sleptons in the MSSM. 
In order to avoid squark and/or charged slepton condensations, 
 which break the gauge symmetries of QCD and/or QED, 
 soft mass squared for squarks and sleptons should be positive. 
If this is the case for the Higgs doublets, 
 the electroweak scale can be stable as follows. 
For large soft masses, 
 we can neglect the D-term potentials in Eq.~(\ref{case 2})  
 and find a potential minimum at 
 $ \langle H_u^0 \rangle \simeq v \sqrt{m_d/m_u}$ 
 and then $\tan \beta \simeq m_d/m_u$. 
Therefore if two soft SUSY breaking masses are of the same order, 
 the electroweak scale is stable 
 even if the soft masses are very heavy. 
As in the case (i), through numerical analysis, 
 we can find that 
 the light Higgs bosons and Higgsinos 
 obtain masses of order $m_{u,d}$. 
Again, all the superpartners and also Higgs bosons 
 can be decoupled without fine-tuning. 
On the other hand, 
 if $m_u^2$ and/or $m_d^2$ are negative 
 the Higgs VEVs are found to be of order $|m_{u,d}|^2$ 
 as in the MSSM. 
Thus the electroweak scale SUSY breaking is necessary 
 and the superpartners cannot be decoupled. 
However, in this case, we find that 
 our model has a similar structure 
 to the recently proposed ``fat Higgs'' model \cite{fat-Higgs}. 
The electroweak symmetry is broken at the SUSY level, 
 and the tree level upper bound on 
 the lightest Higgs boson mass $ \leq M_Z$ in the MSSM 
 can be relaxed. 
In fact, through numerical analysis, 
 we can find that the lightest Higgs boson mass 
 can be $130$ GeV, for example, even at the tree level. 
This phenomenology is worth investigating. 
Case (iii): $|m_{u,d}| \simeq |B_{1,2}|$. 
This case is realized in normal supergravity scenario. 
If both of the soft mass squareds are positive, 
 we can obtain almost the same results in case (ii). 
In other cases, results are depend on input values 
 of the soft SUSY breaking terms. 
More elaborate numerical studies are needed. 
We leave this issue for future works. 

Finally we introduce a model which can naturally realize 
 the superpotential of Eq.~(\ref{superpotential}) 
 as an effective superpotential. 
We present this model as an existence proof of 
 a concrete model rather than a proposal of a specific model. 
One may be able to construct a simpler model 
 than that we will present. 
The most important part in the superpotential is 
 the second term, the runaway type superpotential. 
Although it is hard to belive 
 that any perturbative theories can derive it, 
 such a superpotential in fact can be generated dynamically 
 in SUSY gauge theories \cite{ADS}, 
 where $M$ stands for the dynamical scale 
 of a strong gauge interaction. 
As will be seen in the following, 
 the runaway term can be generated 
 in a composite model of the Higgs doublets. 
However in general it would not be essential 
 for a model to be a composite model. 
Whatever a model is, 
 a quit complicated dynamical model 
 or a model inspired by string theories 
 through some complicated (non-perturbative) structures etc., 
 our aim can be accomplished 
 if only the superpotential of Eq.~(\ref{superpotential}) 
 is finally generated as an effective superpotential. 
We may define the $U(1)_R$ charge for $H_u$ and $H_d$ as $-1/\alpha$ 
 and for the mass term $m$ as $ + 2 (1+1/\alpha)$. 
According to the method in SUSY gauge theories 
 developed by Seiberg and co-workers \cite{Seiberg}, 
 there is a possibility 
 that the superpotential of Eq.~(\ref{superpotential}) 
 can be effectively generated 
 after integrating out other fields in a model, 
 since the superpotential is the unique one 
 consistent with the global $U(1)_R$ symmetry. 

Now we present a model in five dimensions 
 with the warped compactification of the fifth dimension 
 on $S^1/Z_2$ \cite{Randall-Sundrum} with a metric, 
$d s^2 = e^{-2 r_c k |y|} \eta_{\mu \nu} dx^\mu dx^\nu - r_c^2 d y^2$,  
 where $r_c$ and $y$ is the radius and the angle of $S^1$, 
 and $k$ is the AdS curvature scale. 
We follow the formalism in Ref.~\cite{Marti-Pomarol}. 
The model is based on the gauge group 
 $SU(N_c)_H \times SU(2)_L \times U(1)_Y$ 
 with an integer $N_c \geq 3$. 
The particle contents are as follows:  
\begin{center}
\begin{tabular}{c|ccc}
 \hspace{1cm}& $~SU(N_c)_H$~ & $~SU(2)_L~$ & $U(1)_Y$ \\
\hline
$ P $   & $ \bf{N_c^*}$     & $ \bf{2} $  & $  0 $   \\
$ N_u $ & $ \bf{N_c}  $     & $ \bf{1} $  & $+1/2$   \\
$ N_d $ & $ \bf{N_c}  $     & $ \bf{1} $  & $-1/2$   \\ 
\hline  
$S    $ & $ \bf{1}  $   & $ \bf{1} $  & $ 0$   \\ 
$S_c  $ & $ \bf{1}  $   & $ \bf{1} $  & $ 0$   \\ 
$Z    $ & $ \bf{1}  $   & $ \bf{1} $  & $ 0$   \\ 
\end{tabular}
\end{center}
Here $SU(N_c)_H$ is a strong gauge interaction newly introduced. 
Suppose that the singlet superfields ($S$ and $S_c$) reside 
 in the bulk with assigned $Z_2$-parity 
 (even for $S$ and odd for $S_c$), 
 while the singlet ($Z$) and the preons ($P$, $N_u$ and $N_d$) 
 reside on the boundary branes at $y=0$ and $y=\pi$, 
 respectively. 
The basic Lagrangian is given by 
\begin{eqnarray} 
&{\cal L}_{bulk} & =
   \int d^4 \theta r_c \omega(y)^2 
   \left(S^\dagger S + S_c^\dagger S_c \right) \nonumber \\
  &+&   \int d^2 \theta \omega(y)^3 
   S_c \left\{ \partial_y- \left( \frac{3}{2} -c \right) 
   r_c k \epsilon(y) \right\} S +h.c.,  \nonumber \\ 
& {\cal L}_{y=0} & =  
   \int d^4 \theta Z^\dagger Z 
 + \int d^2 \theta 
 Z \left(  \frac{S(0)^2}{M_5}  - M_5^2 \right)  + h.c. , 
 \nonumber \\ 
& {\cal L}_{y=\pi} & = 
   \int d^4 \theta \omega(\pi)^2  K_{preons}  \nonumber \\
 &+&
  \int d^2 \theta \omega(\pi)^3  
  \frac{S(\pi) [P N_u][P N_d]}{M_5^{\frac{5}{2}}} +h.c.,
\end{eqnarray} 
where $M_5$ is the five dimensional Planck scale, 
 $\omega(y)= \exp(-r_c k |y|)$, 
 $c $ is the bulk mass for the bulk fields, 
 $S(y)$ denotes the value of $S$ at $y$, 
 and  $K_{preons}$ denotes the Kahler potential for the preons. 
Normally only $S$ (the $Z_2 $ even field) 
 can couple to the brane fields, 
 and we have introduced the higher dimensional interaction terms 
 naturally suppressed by the five dimensional Planck scale. 

At low energies where $SU(N_c)_H$ becomes strong, 
 the effective Lagrangian is expressed 
 by the effective Higgs fields composed of the preons, 
 $H_u \sim [P N_u]/\Lambda$ and $H_d \sim [P N_d]/\Lambda$ 
 with the $SU(N_c)_H$ dynamical scale $\Lambda$, 
 and new term in the superpotential, 
\begin{eqnarray} 
W_{dyn} = (N_c-2) \left(\frac{\Lambda^{3 N_c-4}}{H_u H_d}
    \right)^{\frac{1}{N_c-2}}  ,
\end{eqnarray} 
is dynamically generated \cite{ADS}. 
For simplicity, we assume the canonical Kahler potentials 
 for the Higgs doublets in the following. 
Then the effective Lagrangian on the boundary brane at $y=\pi$ 
 is rewritten as 
\begin{eqnarray} 
& {\cal L}_{y=\pi} & = 
   \int d^4 \theta \omega(\pi)^2  
 \left( H_u^\dagger H_u + H_d^\dagger H_d  \right)  \\
 &+&
  \int d^2 \theta \omega(\pi)^3  
 \left( 
 \frac{\Lambda^2 S(\pi) (H_u H_d) }{M_5^{\frac{5}{2}}} + W_{dyn}  
  \right) +h.c.,  \nonumber 
\end{eqnarray} 

In order to obtain the 4 dimensional effective Lagrangian, 
 first solve the equation of motion 
 for the bulk field $S$ such as 
 $\left\{ \partial_y -(3/2 -c)r_c k \epsilon(y) \right \} S=0$,  
 and find the solution 
 $ S(x,y) = \tilde{S}(x) \times \exp[(3/2 -c) r_c k |y|]$ 
 with $\tilde{S}(x)$ being 4 dimensional superfield. 
Then substitute the solution into the above Lagrangian, 
 and integrate it with respect to 
 the fifth dimensional coordinate $y$. 
Furthermore, 
 by rescaling and normalizing all the fields appropriately 
 to make their Kahler potentials the canonical forms, 
 we obtain a 4 dimensional effective superpotential, 
\begin{eqnarray} 
 W_{eff} &=& 
 Z \left( \frac{\tilde{S}^2}{N_S^2 M_5} - M_5^2 \right)  
+ 
\omega_c 
 \left( \frac{\Lambda^2}{N_S M_5^{\frac{5}{2}}} \right) 
 \tilde{S} H_u H_d  
 \nonumber \\
& +&  (N_c-2)  \frac{ \left( \omega(\pi) \Lambda 
 \right)^{\frac{3N_c-4}{N_c - 2 }}}
 {(H_u H_d)^{\frac{1}{N_c-2}}} 
\end{eqnarray} 
where $\omega_c= \exp[(1/2 -c) r_c k \pi]$ is defined, 
 and $N_S^2 = [\omega_c^2 -1]/(1-2 c) k$  
 is the normalization factor of $\tilde{S}$. 
After integrating out $\tilde{S}$ and $Z$ 
 under the SUSY vacuum conditions, 
 we finally obtain the effective superpotential 
 of Eq.~(\ref{superpotential}) 
 with the identifications 
\begin{eqnarray} 
\alpha=\frac{1}{N_c-2}, \; \; 
m = \omega_c \frac{\Lambda^2}{M_5}, \; \; 
M = \omega(\pi) \Lambda .
\end{eqnarray} 
We can realize $M \ll m$ with a small warp factor $\omega(\pi) \ll 1$  
 and/or a large $\omega_c \gg 1$. 
Note that $M=\omega(\pi) \Lambda$ is the physical dynamical scale 
 in 4 dimensional effective theory. 
Considering a condition $\Lambda \leq M_5$ 
 for the $SU(N_c)_H$ gauge theory to be well-defined 
 and Eq.~(\ref{vev}), 
 we can obtain the theoretical upper bound on $M$ such as 
\begin{eqnarray}
 M \leq  v  \times \left(\frac{M_5}{v} \omega_c 
 \right)^{\frac{N_c-2}{3 N_c-4}}.  
\end{eqnarray}
Taking, for example, 
 $N_c=4$, $\omega_c \sim 1$ (or $c\sim 1/2$) and 
 $M_5 \sim k \sim M_P \simeq 2.4 \times 10^{18}$ GeV, 
 the reduced Planck mass, we find 
 $M \leq 1500$ TeV 
 and the warp factor $\omega(\pi) \sim 10^{-12}$. 

Some comments are in order. 
Since our model is based on the results in SUSY gauge theories, 
 the scale of the soft SUSY breaking terms 
 should be smaller than the dynamical scale $M$ 
 for the consistency of the model. 
In order to incorporate large Yukawa couplings, 
 our model would be extended. 
A simple way is to introduce 
 a pair of elementary Higgs doublets 
 and mass mixings (of order $m$) 
 among them and the composite Higgs doublets. 
With the elementary Higgs doublets 
 (large) Yukawa couplings can be written as in the MSSM,  
 and fermion masses are generated 
 once the elementary Higgs doublets develop their VEVs. 
We can show that 
 the elementary Higgs bosons obtain the VEVs 
 of the electroweak scale through the mass mixings, 
 even though they are all heavy. 

In summary, we have proposed a simple superpotential 
  for the Higgs doublets, 
 where the Higgs doublets develop their electroweak scale VEV 
 at the SUSY level, 
 while one Higgs chiral multiplet is very heavy. 
In a class of soft SUSY breaking terms, 
 we have shown that 
 the electroweak scale can be stable without fine-tuning 
 so that all the sparticles and Higgs bosons 
 can be decoupled from the low energy theories. 
We have presented a concrete model 
 which can dynamically generate the superpotential. 
Various phenomenological applications of our model 
 are possible according to soft SUSY breaking scales 
 to be concerned. 
For example, in our model, we can realize 
 the recently proposed ``split supersymmetry'' scenario \cite{Split-SUSY}
 without fine-tuning (but there is no light Higgs boson). 
As mentioned above, 
 if we consider the electroweak scale SUSY breaking, 
 our model provides the similar structure 
 to the recently proposed ``fat Higgs'' model. 
These applications are worth investigating. 

We would like to thank Noriaki Kitazawa 
 for useful discussions. 
N.O. would like to thank the Abdus Salam International Centre 
 for Theoretical Physics, Trieste, 
 during the completion of this work. 
This work is supported in part 
 by the Grant-in-Aid for Scientific Research 
% (\#16540258, \#16028214, \#14740164, \#15740164) 
 from the Ministry of Education, Culture, Sports, 
 Science and Technology of Japan. 
%
%
%%%%%%%%%%%%%%%%
%\begin{references}
%

%%%%%%%%%%%%%%%%
%
\end{document}